\documentstyle[12pt,aps]{revtex}%

\begin{document}

\newcommand{\bce}{\begin{center}}
\newcommand{\ece}{\end{center}}
\newcommand{\be}{\begin{equation}}
\newcommand{\ee}{\end{equation}}
\newcommand{\bea}{\vspace{0.25cm}\begin{eqnarray}}
\newcommand{\eea}{\end{eqnarray}}
\def\NCA{{Nuovo Cimento } A }
\def\NIM{{Nucl. Instrum. Methods}}
\def\NPA{{Nucl. Phys.} A }
\def\PLA{{Phys. Lett.}  A }
\def\PRL{{Phys. Rev. Lett.} }
\def\PRA{{Phys. Rev.} A }
\def\PRC{{Phys. Rev.} C }
\def\PRD{{Phys. Rev.} D }
\def\ZPC{{Z. Phys.} C }
\def\ZPA{{Z. Phys.} A }
\def\PTP{{Progr. Th. Phys. }}
\def\LNC{{Lett. al Nuovo Cimento} }

{\bf
New schemes for manipulating quantum states using a Kerr cell  }
\vskip 0.3cm
{Marco Genovese and C.Novero}
\vskip 0.2cm
{\it Istituto Elettrotecnico Nazionale Galileo Ferraris, 
Str. delle Cacce 91,  I-10135 Torino}

\vskip 0.9cm

Recently, Quantum Non Demolition (QND)  detection of a single photon has become possible \cite{SI}.This is due to the
discovery of new materials with very high Kerr coupling as the Quantum Coherent Atomic Systems (QCAS) {\cite{QCAS,SI}}
and the Bose-Einstein condensate {\cite{BEC}}.
These great technical improvements could allow the realisation of small Kerr cells, capable of large phase shift, even with a low-intensity probe.
This recent development has a large relevance for application to quantum information theory and foundations of Quantum Mechanics. For example, it has lead to the proposal of  schemes for complete teleportation 
\ {\cite{TVF}}  and for realising quantum gates {\cite{TVF,Chiara}}. 
In this proceeding we describe some other proposals of application of this device, as the fast modulation of quantum interference  {\cite{nos1}}, the generation of optical Schr\"odinger cats {\cite{nos2}} and of GHZ states and the realisation of translucent eavesdropping.

Let us consider first a scheme for realising a fast modulation of quantum interference. A "signal" photon enters a Mach-Zender interferometer through  a Beam-Splitter (BS I) from port 1;
assuming a 50\% BS (the treatment of  the non 50\% case is a trivial extension) one has after the BS

\be
a_2={a_1 + i a_0 \over \sqrt{2} } \, \, \, \, \, a_3={a_0 + i a_1 \over \sqrt{2}}
\label{eq:a2a3}
\ee
A probe laser crosses the Kerr cell on the third arm acquiring a phase which in principle will be measurable, providing { \it welcher Weg} information.
The usual treatment of Kerr interaction gives 
after recombination on Beam Splitter II:
\be
\langle \Psi \vert n_4 \vert \Psi \rangle = {1 
 \over 2} \left [ 1 - \exp [-2 \vert \nu \vert ^2 \sin ^2( \chi T)] 
 \cos [(\omega_s + \chi_s  ) T + \Theta + \vert \nu \vert ^2 \sin (2 \chi T)
] \right ]
\label{eq:n5}
\ee
where $\Theta$ takes into account different lengths of arms 2 and 3 and could be varied interposing a variable phase shift on one of the interferometer arms. $\chi_s$ and $\chi$ denote the self- and the cross- Kerr couplings respectively and $n_3$ and $n_p$ are the photon number operators on the arm 3 and for the probe, which is assumed to be described by a coherent field $\vert \nu \rangle$.

To this point, the usual treatment {\cite{SI}} of QND ideal { \it welcher Weg} experiment has been considered,
with the well known result that the probe laser acquires a phase which permits, by a homodyne measurement, the identification of the path followed by the signal photon. The signal-to-noise ratio in the homodyne measurement, given by $R =4
 \vert \nu \vert \sin (\chi T)$, is directly related to the 
suppression of interference, whose visibility is  $\exp{ ( - R^2 /8)}$. 

Let us now consider the insertion of a second Kerr cell on the $2^{nd}$ arm  and let us assume that the interaction time between the probe and signal fields in this cell is $T'$.
We now have:

\be
\langle \Psi \vert n_4 \vert \Psi \rangle 
= {1 \over 4 } \langle \Psi \vert n_1 [ 2 - ( \exp {[-i \left ( \Theta +  \beta ( T-  T' ) \right ) ]} +
\exp {[i \left (\Theta + \beta ( T -  T' ) \right ) ]})] \vert \Psi \rangle
\ee
where $\beta = (\omega_s + \chi_s /2 + \chi_s n_s + 2 \chi n_p )$.
If we choose the Kerr cells so that $T= T'$, the phase into the probe due to the photon in path 3 or 2 would be the same and the interference pattern 
${1 - \cos (\Theta) \over 2}$ is recovered for the signal field.
On the other hand, if one considers the case where the distance between
the two Kerr cells is larger than the coherence length of the probe laser the two paths will still be distinguishable  and interference will be lost.
Quantum interference can thus  be regulated, changing the coherence length of the laser before injecting it into the first Kerr cell. The observation of this effect represents a very good and illustrative  example of the effect of disappearance of quantum interference when {\it welcher Weg} information is obtained and of the effect of erasing this information.

Let us now substitute the first beam splitter with a Polarising-Beam-Splitter I (PBS I) and let us suppose that the entering photon is in a superposition of vertical (V) and horizontal (H) polarisation, which will take different paths, for example the vertical one will follow path 2 and the horizontal path 3. 
As before, a probe laser crosses the Kerr cell on the  arm 3 acquiring a phase or not according if the photon crosses or not the cell.
Thus the entangled state:
\be
\vert \Psi \rangle = { \vert H \rangle  \vert \nu '\rangle + \vert V \rangle  \vert \nu \rangle \over \sqrt {2}}
\label{eq:Psi}
\ee
is generated, where the coherent state $\vert \nu '\rangle$ differs in phase from $\vert \nu \rangle$.\newpage
The two
 signal photon paths are then recombined on a second beam splitter
 (BSII) and a polarisation measurement is performed on this photon on the base at $45 ^o$.  This is the conditional measurement producing the Schr\"odinger cat: if the signal photon is found to have a  $45 ^o$ ($135^o$) polarisation, the coherent state is projected into the superposition
\be
\vert \psi_{+ (-)} \rangle = {  \vert \nu '\rangle + (-)   \vert \nu \rangle \over \sqrt {2}}
\ee
A superposition of two "many photons"  states is thus obtained.

The one photon signal state can be easily produced, for example,  using parametric down conversion (PDC) in a non-linear crystal. In this case the second photon of the down-converted pair can be used as trigger. Furthermore, if the same pulsed laser is used both for pumping the crystal and for the Kerr effect, one can easily obtain a good timing  for the crossing  of the Kerr cell for the signal photon and the coherent state.
A main advantage of the present configuration  respect to the Schr\"odinger cat generation using cavities is  that the coherent states superposition travels in  air between the Kerr cell and the detection apparatus allowing a much longer time before decoherence takes over.

For identifying the macroscopic superposition one can look for a negative part of the Wigner function, reconstructed by tomographic techniques { \cite{tomo}}.We have performed a numerical simulation  keeping  into account the deterioration due to errors: our results show that  even with $25 \%$, or larger, errors on the homodyne measurement  the cat can be easily identified.

The use as input of a photon from PDC in a Kerr cell allows also the creation of a three photons GHZ entangled state, which is one of the three elements necessary for realising an optical quantum computer  {\cite{QC}}, together with single qubit operations, which are easily implemented, and teleportation. For what concerns this last  , a description of a scheme performing this operation using a Kerr cell appears in Ref.  {\cite{TVF}}. Using the same polarisation dependence of the Kerr interaction of Ref.  {\cite{TVF}}, where there is no effect except when both the photons interacting in the Kerr cell have vertical polarization ($\vert V \rangle \vert V \rangle \rightarrow \vert V \rangle \vert V \rangle e^{i \phi}$), a GHZ state can be generated by the interaction of an entangled pair of photons with a third one.
Let us assume, for example, of having generated the entangled state {\cite{nos3}}:
$\vert \Phi ^+ \rangle =  \vert H \rangle \vert H \rangle + \vert V \rangle 
\vert V \rangle / \sqrt{2}$.
A simple calculation shows that  the interaction in the Kerr medium of the second photon of the pair with a third photon polarised at $45 ^o$ (denoted by $\vert 45 \rangle$, whilst its orthogonal state is $\vert 135 \rangle$), is, for a phase shift $\phi= \pi / 2$,  the GHZ state:
\be
\vert \Psi _{GHZ} \rangle = { \vert H \rangle \vert H \rangle \vert 45 \rangle + \vert V \rangle 
\vert V \rangle \vert 135 \rangle \over \sqrt {2}}.
\label{eq:GHZ}
\ee 
Analogous results are easily derived for the other three Bell states.

Finally, let us notice that such an apparatus can also be used for performing translucent eavesdropping on a quantum channel where polarisation is used for distinguishing qubits. In this case the input port of the first beam splitter is fed with a single photon of vertical polarisation, which splits on the two interferometer arms. On arm 1 it interacts with the transmitted qubit inside the Kerr cell (with the same polarisation dependence as before).
Let us suppose that the transmitted qubit is in the general form
$\vert u \rangle = cos (\theta) \vert H \rangle + sin( \theta ) \vert V \rangle
$, denoting the probe photon with $ \vert p \rangle$, the final state is:
\be
\vert \Psi \rangle = { 1 \over \sqrt{2}} \left [cos (\theta) \vert H \rangle \vert p \rangle_1 + sin( \theta ) \vert V \rangle \vert p \rangle_1 e^{(i \phi)}+
i \vert u \rangle \vert p \rangle_2 \right ]
\ee
where the suffixes after the probe $\vert p \rangle$ denotes the path followed and where we have considered  a 50 \%  BS (which gives Eve the largest information on Alice-Bob transmission).
We have thus obtained the desired entanglement between the probe and the transmitted qubit, which can be used for  translucent or coherent eavesdropping.

\end{document}